\documentstyle[12pt,fleqn,cite,epsfig]{article}

\def\be{\begin{equation}}
\def\ee{\end{equation}}
\def\bea{\begin{eqnarray}}
\def\eea{\end{eqnarray}}

\begin{document}
\begin{titlepage}
\begin{center}
\hfill hep-th/0102138\\
\hfill IP/BBSR/2001-03\\
\vskip .2in

{\Large \bf Holography and Stiff-matter on the Brane}
\vskip .5in

{\bf Anindya K. Biswas\footnote{e-mail: anindya@iopb.res.in} and Sudipta
Mukherji\footnote{e-mail: mukherji@iopb.res.in}\\
\vskip .1in
{\em Institute of Physics,\\
Bhubaneswar 751 005, INDIA}}

\end{center}

\begin{center} {\bf ABSTRACT}
\end{center}
\begin{quotation}\noindent
\baselineskip 10pt
Recently, Verlinde noted a surprising similarity between Friedmann 
equation
governing radiation dominated universe and Cardy's entropy formula in
conformal field theory. In this note, we study a brane-universe filled
with
radiation and stiff-matter. We analyze Friedmann equation in this context
and compare our results with Cardy's entropy formula.

\end{quotation}
\vskip 2in
February 2001\\
\end{titlepage}
\vfill
\eject

Since 't Hooft \cite{GT} and Susskind \cite{LS} proposed the idea of
holography, there have been
several works in the past trying to test this proposal in various gravity
systems. 
According to the holographic principle, dynamics of $D$ space-time
dimensions are encoded in terms of the degrees of freedom on the $D-2$
dimensional boundary. The ADS/CFT correspondence does indeed realize such
a principle. In systems where gravity is dynamical, testing holographic 
principle, however, turned out to be somewhat subtle. Fischler and
Susskind \cite{Fish}, in particular, addressed the issue of holographic
entropy bound
in the cosmological setting. They proposed that the area of particle
horizon should bound the entropy on the backward looking
light-cone. However, violation of such bound was found in case of closed
FRW universe. Subsequently, various modifications \cite{VA} were proposed
to circumvent such a problem. 

In a more recent paper \cite{EV}, Verlinde analyzed entropy bound in a
radiation dominated closed FRW universe using the holographic principle.
Representing radiation as conformal field theory (CFT) in the universe, he
found surprising similarity between Cardy's entropy formula for $1 + 1$ 
dimensional CFT and the Friedmann equation governing the evolution of the
universe. In fact, it turned out after suitable identifications, that 
Cardy's formula maps to Friedmann equation. Furthermore, in \cite{EV},
a cosmological entropy bound is proposed which 
unites Beckenstein bound \cite{JB} for limited self-gravity system and
Hubble
bound \cite{Ven} for strong self-gravity system in an elegant
way.\footnote{See also \cite{various} for further work on analyzing  
Verlinde's proposal in different circumstances.} 
Though, Verlinde's analysis was for
radiation dominated universe (which can be modeled perhaps by CFT), the
arguments 
of the paper, leading to entropy bound, depend on the equation of state of
matter in a very minimal way. This might lead one to suspect that a
version of Verlinde's bound might hold even when the universe is described
by 
a different equation of state. The purpose of this note is to test
Verlinde's proposal for an universe filled with radiation and stiff
matter.
Following closely \cite{SV}, we consider a brane-world in a charged 
black
hole background \cite{CEJM}. The time development of the metric on the
brane is described by Friedmann equation \cite{SG,SV}. The mass of the
black hole background 
induces radiation matter on the brane \cite{SV}. On the other hand, the
charge of the black hole, however, acts as a source for stiff matter on
the
brane ( whose equation of state is simply given by the pressure being
equal to the 
energy density). We show that when the brane crosses the horizon of the
charged  black hole, entropy and temperature can be expressed in terms
of the Hubble parameter and its derivative. In particular, none of these
thermodynamical quantities depend explicitly on mass and charge of the
background. This is similar to what was observed in \cite{SV}, when 
the target space was uncharged and hence the brane is described by 
radiation 
dominated matter only. We also, some what surprisingly, find that the
formula
for entropy given in \cite{SV}, from the CFT consideration, coincides with
the FRW equation when the brane crosses the horizon of the charged black
hole.

We start by considering an $(n+1)$ dimensional brane in a space-time
described by an (n+2) dimensional charged AdS black hole. The background
metric is thus given by \cite{CEJM}
\begin{equation}
ds_{(n+2)}^2 = - h(a) dt^2 + {{da^2}\over{h(a)}} + a^2 d\Omega_{n}^2,
\label{metric}
\end{equation}
where 
\begin{equation}
h(a) = 1 - {\omega_{n+1} M\over{a^{n-1}}} + {n \omega_{n+1}^2 Q^2
\over {8 (n-1) a^{2n-2}}} + {a^2\over{L^2}}.
\label{component}
\end{equation}
Here, $M$ and $Q$ are respectively ADM mass and charge, $L$ is the
curvature radius of  space and $\omega_{n+1} 
={16 \pi {\bf G}_N/(n {\rm Vol}(S^n))}$. ${\bf G_N}$ here is the Newton's
constant. 
If $a_H$ is the largest real positive root of $h(a)$, then in order for
(\ref{metric}) to describe a charged black hole with non singular horizon
at $ a = a_H$, the later should satisfy
\begin{equation}
\Big({n\over{n-1}}\Big) a_H^{2n} + L^2 a_H^{2n -2} \ge {nL^2
\omega_{n+1}^2
Q^2\over {8(n-1)}}.
\end{equation}
The electrostatic potential difference between the horizon and infinity is
given by the quantity ${\Phi}$ and following \cite{CEJM} we choose it
to be 
\begin{equation}
{ \Phi} = \Big({n\over{4n-4}}\Big){\omega_{n+1} Q\over{a^{n-1}}}.
\label{potential}
\end{equation} 
This space-time was analyzed with a focus  on  AdS/CFT correspondence in
\cite{CEJM}. 
\footnote{Interestingly enough, it was found that in the high temperature
limit the
charge essentially
plays no role (see the discussion in the last section of \cite{CEJM}). So
at least in this limit the CFT that lives in the
boundary is expected to behave in the same manner as was discussed by
Witten \cite{EW}.} 

Let us now consider a $(n+1)$-dimensional brane with a constant tension in
the background of (\ref{metric}). As discussed in \cite{SV}, if we regard
the brane as a boundary of background AdS geometry, location and the
metric on the boundary becomes time dependent. The induced equation of
motion of the metric on the brane satisfy Friedmann equation.\footnote{
The way to arrive at such equation has been discussed in
\cite{MV,SG,CEG,SV},
we will skip the detail here.}
 
Tuning the brane
cosmological constant to zero, we get the following induced equation on
the brane:
\begin{equation}
H^2 = -{1\over {a^2}} + {\omega_{n+1} M\over {a^{n+1}}} - 
{ n \omega_{n+1}^2 Q^2\over{8 (n-1) a^{2n}}},
\label{freedman}
\end{equation}
where $H = {\dot a\over a}$ being the Hubble parameter and dot denotes
derivation with respect to the brane time $\tau$. Someone who is living
on the brane will thus observe a closed FRW metric given by
\begin{equation}
ds_{n+1}^2 = -d\tau^2 + a(\tau)^2 d\Omega_n^2,
\end{equation}
with $a(\tau)$ satisfying (\ref{freedman}). Now  from different
power law behaviour of the quantities in the right hand side of
(\ref{freedman}) with respect to  
$a$, we conclude that
while $M$ behaves as a source of radiation matter as in \cite{SV}, $Q$
behaves as a source of stiff matter on the brane.
However, we should notice that the $Q^2$ dependent term in
(\ref{freedman}) has appeared with a wrong sign in front. 
Differentiating (\ref{freedman}) with respect to $\tau$ we get
\begin{equation}
\dot H =  {1\over{a^2}} - {(n+1)\omega_{n+1} M\over{2 a^{n+1}}}
+ {n^2 \omega_{n+1}^2 Q^2\over{8(n-1)a^{2n}}}.
\label{hdervetive}
\end{equation}

At this point, we would like to comment on the cosmological behaviour
of the universe on the brane. The time development of the brane-universe
is completely determined by (\ref{freedman}). Like closed
model for radiation dominated universe, brane-universe reaches a maximum
value $a_{\rm max}$ which is given by the largest real root of the
equation, 
\begin{equation}
{1\over {a^2}} + {\omega_{n+1} M\over {a^{n+1}}} -
{ n \omega_{n+1}^2 Q^2\over{8 (n-1) a^{2n}}} = 0.
\label{max}
\end{equation}
Furthermore, it is easy to check that there is a non-zero minimum
value for the scale
factor $a_{\rm min}$. So the brane-universe oscillates between 
$a_{\rm min}$ and $a_{\rm max}$. It can also easily be seen that 
$a_{\rm min} < a_H < a_{\rm max}$, where $a_H$ is the location of the
horizon of the black hole background given in (\ref{metric}).
Now, if we compare (\ref{max}) with (\ref{freedman}), we see that the
Hubble parameter satisfy $H^2 = {1\over {L^2}}$. This is similar to what
was found in \cite{SV}. We also note that $H$ 
is independent of black hole parameters $M$ and $Q$ at the horizon.

If we take the brane as the boundary of the AdS space, following
\cite{EW}, one would expect that the entropy of the CFT is given by the 
Bekenstein-Hawking entropy of the charged AdS black hole
\begin{equation}
S = {{a_H^n} {\rm Vol}(S^n)\over{4 {\bf G}_N}}.
\end{equation}
From above, it follows that the entropy density on the brane is 
\begin{equation} 
s = (n-1) {a_H^n\over {4 G_N L a^n}},~~{\rm and ~at} ~a = a_H, 
~s = {(n-1)H\over {4G_N}},
\label{braneentropy}
\end{equation}
where $G_N$ is the Newton's constant on the brane. It is related to ${\bf
G}_N$ via $G_N L = (n-1) {\bf G}_N$.
These expressions are the same as that obtained in \cite{SV} where,
the bulk space-time was an uncharged black hole. 
These expression are some what universal as they do not explicitly depend
upon black hole parameters like mass and charge etc.
Similarly, from the expression of the Hawking temperature for the metric 
in (\ref{metric}), one can obtain the induced temperature on the brane,
which  again does not depend explicitly on black hole parameters at $a
= a_H$ and can be expressed as:
\begin{equation}
T = -{\dot H\over {2 \pi H}}|_{a = a_H}\Big({L\over a}\Big).
\end{equation}
When the brane crosses the horizon, we notice that all the thermodynamical
quantities on the brane can be written down
solely in terms of $H$ and $\dot H$.

The first law of thermodynamics, written in terms of densities, takes the
following form on the brane:
\begin{equation}
T ds = d\rho - \Phi d\tilde \rho + n ( \rho + p -  \Phi \tilde \rho -
Ts){da\over a}.
\label{thermo}
\end{equation}
Here, $\rho$ is the energy density, $p$ is the pressure, and $\tilde \rho$
is the charge density. The quantity $(\rho + p - \Phi \tilde \rho - Ts)$
is a measure of non-extensiveness of thermodynamical quantities. Now using
the identification,
\begin{eqnarray}
\rho &=& {ML\over {a^{n+1}{\rm Vol}(S^n)}},~~
p = {ML\over {n a^{n+1}{\rm Vol}(S^n)}},\\
\Phi &=& \Big({n\over{4n -4}}\Big){\omega_{n+1} Q L\over{a^{n}}},~~
\tilde \rho = {Q\over{a^{n}{\rm Vol}(S^n)}},
\end{eqnarray}
we get 
\begin{equation}
{n\over 2}(\rho + p - {\Phi \tilde \rho} - Ts)  = {\gamma\over
{a^2}}.
\label{thermot}
\end{equation}
where 
\begin{equation}
\gamma = {n(n-1)\over{16 \pi G_N}} {a_H^{n-1} \over {a^{n-1}}}.
\end{equation}

Following the same line arguments as \cite{SV}, we can express the entropy
density in the form of Cardy's formula
\begin{equation}
s = \Big({4\pi\over{n}}\Big){\sqrt{\gamma\Big(\rho - {\Phi\tilde
\rho\over 2}
-{\gamma\over {a^2}}\Big)}}.
\label{cardy}
\end{equation}
It is easy now to check that the equations (\ref{cardy}) and
(\ref{thermot}) are the same as Friedmann equations (\ref{freedman})
and (\ref{hdervetive}) at $a = a_H$. When expressed in terms of total
entropy, (\ref{cardy}) reduces to the one discussed in \cite{EV}.

Appearance of the quantity $\Phi\tilde\rho$ in (\ref{thermo}), 
(\ref{thermot}) and (\ref{cardy}) is interesting. From the point of view
of background black hole geometry, it measures the energy density due to 
charge. On the other hand, we expect that from the brane perspective, it 
should be the contribution of the energy density and pressure (which are
equal in magnitude however) of stiff matter. It is satisfying to observe
that 
\begin{equation}
\Phi \tilde \rho = \rho_{\rm sm} + p_{\rm sm} ~{\rm at} ~a= a_H,
\end{equation}
where subscripts on $\rho$ and $p$ indicate that they are stiff-matter
contributions. We would like to emphasise that it is at $a =a_H$, where
the entropy formula (\ref{cardy}) and (\ref{thermot}) 
coincides with Friedmann equation (\ref{freedman}) and (\ref{hdervetive})
respectively.

\vskip .3in

\noindent {\bf Acknowledgements}

\noindent We would like to thank A. Kumar, B. Rai and A. Srivastava for
general discussions on the holographic principle. SM would like to thank
P. Parashar for her support during the course of this work.

\vfil
\eject


\end{document}